\documentclass[final,5p,times,twocolumn]{elsarticle}
\usepackage{graphicx}
\usepackage{amssymb}
\usepackage{lineno}
\usepackage{color}   
\usepackage{amsmath}
\usepackage{amssymb}
\usepackage{comment}
\usepackage{caption}
\usepackage{subfig}  
\usepackage{amsfonts}

\journal{Neurocomputing}

\begin{document}
\begin{frontmatter}
\title{Symbolic analysis of bursting dynamical regimes of Rulkov neural networks}
\author[ufpr]{R. C. Budzinski}
\author[ufpr]{S. R. Lopes}
\author[upc]{C. Masoller}\corref{cor1}
\ead{cristina.masoller@upc.edu}

\address[ufpr]{Department of Physics, Universidade Federal do Paran\'a, 81531-980 Curitiba, Paran\'{a}, Brazil.}
\address[upc]{Department of Physics, Universitat Politecnica de Catalunya, Rambla St. Nebridi 22, 08222 Terrasa, Barcelona, Spain.}

\begin{abstract}
Neurons modeled by the Rulkov map display a variety of dynamic regimes that include tonic spikes and chaotic bursting. Here we study an ensemble of bursting neurons coupled with the  Watts-Strogatz small-world topology. We characterize the sequences of bursts using the symbolic method of time-series analysis known as ordinal analysis, which detects nonlinear temporal correlations. We show that the probabilities of the different symbols distinguish different dynamical regimes, which depend on the coupling strength and the network topology. These regimes have different spatio-temporal properties that can be visualized with raster plots. 
\end{abstract}
\end{frontmatter}

\section{Introduction}\label{1}

Neurons encode and transmit information in temporally correlated sequences of spikes \cite{book_neural_code,kandel2000,quiroga2009extracting}. Neurons can fire regularly (tonic spikes) or irregularly. In the first case, the spikes are periodic in time and the distribution of inter-spike intervals is very narrow; in the second case, the time intervals between spikes are irregular and their distribution is broad. Bursting is a dynamical regime in which a neuron fires groups or bursts of spikes and each burst is followed by a silent period before the next burst occurs.

Since spike correlations have a functional role in the neural code, improving or degrading information transmission~\cite{nelson,andre,andre2,neural_correl}, relevant questions are how to detect temporal correlations in the spike sequences and how the couplings among the neurons affect them. While linear correlations can be detected and quantified by serial correlation coefficients~\cite{neiman_pre_2005,nawrot2007serial,lindner_2013,braun2017evolution}, nonlinear correlations need to be detected by using nonlinear techniques~\cite{book_nltsa}. A popular one is symbolic ordinal analysis \cite{bandt2002permutation,book_amigo}, which applied to sequences of spikes detects nonlinear temporal correlations \cite{rosso_2009,li_2011,parlitz2012classifying,zanin2012permutation,rubido2011language,reinoso2016analysis,masoliver_2018,masoliver_2019,masoliver_2020,maria_preprint}. 

When used to analyze burst sequences, this symbolic approach considers only the relative duration of the time intervals between bursts (inter-burst-intervals, $IBI$s), and transforms a sequence of $IBI$s into a sequence of symbols (known as ordinal patterns) using the ordinal rule that takes into account the temporal order of consecutive time intervals. For example, when analyzing the symbols defined by three consecutive intervals, 
the six possible order relations define six symbols whose 
frequencies of occurrence can reveal the presence of temporal order, in the form of over-expressed and/or less-expressed symbols. If the burst sequence is fully stochastic the frequency of occurrence of each symbol will be $\sim 1/6$ (if the sequence is long enough). On the other hand, over (or less) expressed symbols (whose probabilities are significantly higher or lower than $1/6$) unveil the presence of temporal correlations among consecutive bursts. 

Here we use ordinal analysis to investigate nonlinear correlations in a neuronal ensemble modeled by the popular Rulkov two-dimensional iterated map \cite{rulkov2001regularization}. We use this map because it displays a variety of dynamical regimes and makes possible to simulate the behavior of relatively large neuronal ensembles \cite{rulkov2002modeling}. We focus on the regime where individual neurons fire bursts of spikes and analyze temporal correlations among the $IBI$s. We find that ordinal analysis detects correlations in the $IBI$ sequences that depend on the strength of the coupling, $\varepsilon$, between neurons and on the network topology that is varied from regular to small-world by changing the rewiring parameter, $p$, defined Watts and Strogatz~\cite{watts_1998}. The analysis of the ordinal probabilities as a function of $\varepsilon$ and $p$ reveals different dynamical regimes, which can be clearly visualized using network spatio-temporal plots (raster plots). 

Complementing the symbolic analysis, we also use the Kuramoto order parameter \cite{kuramoto2012chemical} to investigate the synchronization features of the network in the parameter space of $\varepsilon \times p$, in which the role of the coupling and topology are considered. We find that the network displays a variety of complex spatio-temporal patterns, including phase-synchronized states~\cite{yu2011chaotic, hong2002synchronization} and un-synchronized states where zig-zag structures are seen in the raster plots \cite{wang2008synchronization,budzinski2019synchronous,osipov2005synchronized,ivanchenko2007network}.

The paper is organized as follows: Sec.~\ref{2} describes the neuron model and the network topology; Sec.~\ref{3} describes the methodology used to analyze the network dynamics; Sec.~\ref{4} presents the results, Sec.~\ref{5} presents the discussion, and Sec.~\ref{6} presents the conclusions.

\section{Model}\label{2}

The Rulkov two-dimensional iterated map \cite{rulkov2001regularization,ibarz2011map} is used to simulate an ensemble of $N$ neurons. The model equations are: 
\begin{equation}
x_{t+1,i} = \frac{\alpha_{i}}{1+x_{t,i}^{2}}+y_{t,i} + I_{t,i,\mathrm{coupling}} + I_{i,\mathrm{noise}} \label{rulkov_1}
\end{equation}
\begin{equation}
y_{t+1,i} = y_{t,i} - \beta  (x_{t,i} - 1), \label{rulkov_2}
\end{equation}
where $x_{t,i}$ and $y_{t,i}$ are the fast and slow variables respectively of $i$th neuron at the discrete time $t$. Different combinations of the parameters result in different dynamical behaviours. Here we chose parameters such that  the individual neurons are in the bursting regime~\cite{rulkov2001regularization} (all parameter values are listed in Sec.~\ref{implementation}). 

The coupling term, $I_{t,i,\mathrm{coupling}}$, models the contribution of other neurons to neuron $i$ at discrete time $t$:
\begin{equation}
I_{t,i, \mathrm{coupling}} = \frac{ \varepsilon}{\mathrm{\chi}}\sum \limits^{N}_{j=1} a_{i,j}x_{t,j},
\label{rulkov_coupling}
\end{equation}
where $\varepsilon$ is the coupling strength, $\chi$ is a normalization factor equal to the average number of connections per neuron (i.e., the total number of connections in the network divided by the number of neurons, $N$), and $a_{i,j}$ is a symmetric adjacency matrix: $a_{i,j} =a_{j,i}= 1 \, (0)$ if neurons $i$ and $j$ are (are not) connected. 

The term $I_{i,\mathrm{noise}}$ represents neural noise, which is uncorrelated for each neuron. 

\begin{figure}[t]
    \centering
    \includegraphics[width=0.95\columnwidth]{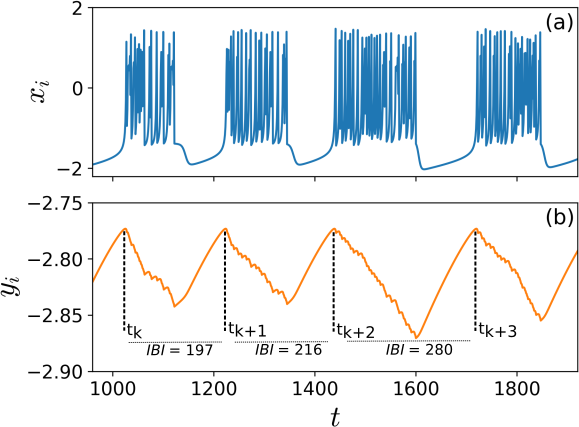}
    \caption{Dynamical behavior of an isolated neuron ($\varepsilon=0$). (a) Fast variable and (b) slow variable of the Rulkov map. The blue line ($x$) depicts the bursting activity where the maxima of $y$ coincide with the bursts beginning. Here $\alpha = 4.25$; similar behavior occurs for other values of $\alpha \in [4.1, \, 4.4]$, other parameters are listed in Sec.~\ref{implementation}.}
    \label{rulkov_dynamics}
\end{figure}

The typical behaviour of an isolated neuron ($\varepsilon=0$) is depicted in Fig. \ref{rulkov_dynamics}. Panel (a) shows the fast variable $x$, which exhibits the burst activity and panel (b) depicts the slow variable $y$, whose maximums coincide with the start of the bursts. 

\section{Methods}\label{3}

\subsection{{Network construction}}

The adjacency matrix of the network, $a_{i,j}$, is constructed following the Watts-Strogatz procedure \cite{watts_1998}, where a network, initially regular, is gradually modified to a small-world one with random characteristics by replacing local connections for random ones. We start with a regular network and, as the re-wiring parameter, $p$, increases, the local connections are replaced by random ones, which leads to the decrease of the average path length without change of the total number of connections. In this way, a small-world network is obtained, where the average path length is low and the clustering coefficient is high \cite{watts_1998}. 

\subsection{{Inter-burst intervals}}

As depicted in Fig.~\ref{rulkov_dynamics}, the maximums of the slow variable $y_i$ allow to detect the times when the bursts of neuron $i$ start. Specifically, the $k$th maximum of $y_i$ that occurs at time $t_{k,i}$ indicates the start of the $k$th burst of neuron $i$.

The $k$th inter-burst interval ($IBI$) of the $i$th neuron is the time interval between two consecutive bursts:
\begin{equation}
    IBI_{\mathrm{k},i} = t_{\mathrm{k+1},i} - t_{\mathrm{k}}.
    \label{ibi}
\end{equation}

The mean value of the inter-burst interval, $\langle \overline{IBI} \rangle$, is calculated by performing a temporal average over the $IBI$ sequence of each neuron,  $\langle {IBI}_i \rangle=\langle IBI_{\mathrm{k},i}\rangle_k$, followed by an average over all neurons: $\langle \overline{IBI} \rangle=\langle IBI_{\mathrm{k},i}\rangle_{\mathrm{k},i}$.

\subsection{Ordinal analysis}\label{3.3}

After calculating the $IBI$ sequence of each neuron, $\{IBI_{\mathrm{k},i}\}$, we use \emph{ordinal analysis} \cite{bandt2002permutation} to detect temporal structures in the $IBI$ sequences. As discussed in the Introduction, ordinal analysis allows identifying patterns in complex datasets. This method takes into account the relative temporal ordering of the data values in a time-series and provides a way to compute a set of probabilities that characterize the time series. 

For example, if we consider two consecutive $IBI$ intervals, if the first one is longer than the next one we assign to that $IBI$ the symbol ``10'', otherwise we assign the symbol ``01''. In this way, we have reduced the sequence of $IBI$ time intervals to a sequence of two symbols. If we compare the relative ordering of three (or four) consecutive values of $IBI$, then six (or twenty-four) different symbols (known as \emph{ordinal patterns}) can be defined. For example, three consecutive increasingly long $IBI$s give pattern $012$, while three consecutive increasingly short $IBI$s give pattern $210$. In this sense, an example of the pattern $012$ is observed in Fig.~\ref{rulkov_dynamics} where a sequence of three increasing $IBI$ values ($197, 216, 280$) is depicted. 

The frequency of occurrence of the different patterns (evaluated from the sequences of IBIs of all the neurons, $\{IBI_{\mathrm{k},i}\}$) give the set of ordinal probabilities, $\mathcal{P}(i)$ with $\sum_i\mathcal{P}(i)=1$, that will be analyzed in the next section.

We consider patterns of length $K=3$. As the number $\mathcal{N}$ of patterns grows as $\mathcal{N}=K!$, using a larger value of $K$ means that there is a large number of probabilities to calculate, which is computationally expensive (see~\cite{reinoso2016analysis} for a discussion of the data requirements). Ordinal patterns of length $K=3$ are small-size patterns, but nevertheless, they yield relevant information about a system's dynamics (see Ref.~\cite{bandt} for practical examples of how small ordinal patterns of length $K=2$ or 3 are able to extract relevant information from data).

It has been recently shown that when a time series contains a significant number of equal values (in our case, equal $IBI$s), they can give rise to false conclusions regarding the presence of temporal structures~\cite{olivares}. Because we chose parameters such that the neurons' dynamics is chaotic, the $IBI$ sequences do not have a large number of equal values (we have verified that the percentage of patterns that contain equal $IBI$s is less than 6\%).

Shannon entropy computed from ordinal probabilities is known as \emph{Permutation Entropy}~\cite{bandt2002permutation}, $S = - {\sum_{i=1}^{\mathcal{N}}\mathcal{P}(i)\ln{(\mathcal{P}(i))}}$. $S$ is maximum when $\mathcal{P}(i)=1/\mathcal{N}$ $\forall$ $i$ and $S=0$ when $\mathcal{P}(i)=1$, $\mathcal{P}(j)=0$ $\forall$ $j\ne i$.

\subsection{Phase synchronization quantifier}\label{3.1}

We use the Kuramoto order parameter~\cite{kuramoto2012chemical} to measure the degree of phase synchronization of the bursts of the neuronal ensemble. The Kuramoto order parameter is given by
\begin{equation}
R(t)=\left|\frac{1}{N}\sum_{j=1}^{N}\mathrm{e}^{i\theta_{j}(t)}\right|,
\label{order_parameter}
\end{equation} 
where $\theta_{j}(t)$ is the phase of neuron $j$ at time $t$. The time-averaged parameter, $\langle R \rangle_t$, is $\sim 1$ if the neurons are phase-synchronized and is $\sim 0$ if they are phase incoherent. 

To associate a phase to the dynamics of each neuron, we use the slow variable, $y$. As depicted in Fig. \ref{rulkov_dynamics}, $y$ is maximum when a burst starts, and therefore, the phase as a function of time can be defined as~\cite{ivanchenko_2004}:
\begin{equation} 
\theta_{i}(t)=2\pi \mathrm{k} +2\pi\frac{t-t_{\mathrm{k},i}}{t_{\mathrm{k}+1, i}-t_{\mathrm{k},i}}, 
\hspace{0.5cm} t_{\mathrm{k},i}<t<t_{\mathrm{k}+1,i},
\label{phase}
\end{equation}
where $t_{\mathrm{k},i}$ is the time where the $k$th burst of the $i$th neuron starts. The parameters considered in this study are such that all neurons show bursting behavior, which allows the use of $y$ maximums to detect the beginning of the bursts, which in turn allows defining a phase without the need of reconstructing the dynamical evolution in the phase space \cite{boccaletti_2002}.

\subsection{Implementation}\label{implementation}

The model parameters are: $\beta= 0.001$ (equal for all neurons) and $\alpha_{i} \in [4.1, \, 4.4]$ Gaussian distributed with mean value $4.25$ and standard deviation $0.045$. $I_{i,\mathrm{noise}}$ is uncorrelated for each neuron: $I_{i,\mathrm{noise}} \in [0.003, \, 0.065]$ is Gaussian distributed with mean value $0.035$ and standard deviation $0.01$. The coupling strength and the rewiring probability are considered control parameters, varied in the range $\varepsilon \in [0.000,\,0.099]$ and $p \in [0.001,\,1.000]$.

The initial conditions are random in the interval $x,\, y \in [0,1]$.

The Kuramoto parameter, $\langle R \rangle_t$, the mean IBI, $\langle \overline{IBI} \rangle$, and the ordinal probabilities were calculated by averaging results from $10$ simulations with different networks, different parameters and different initial conditions.

To investigate the role of the system size, we have simulated networks of 100, 500, and 1000 neurons with 400, 2000, and 4000 connections, respectively, which give the same average of four connections per neuron.

\begin{figure}[t]
    \centering
    \includegraphics[width=0.9\columnwidth]{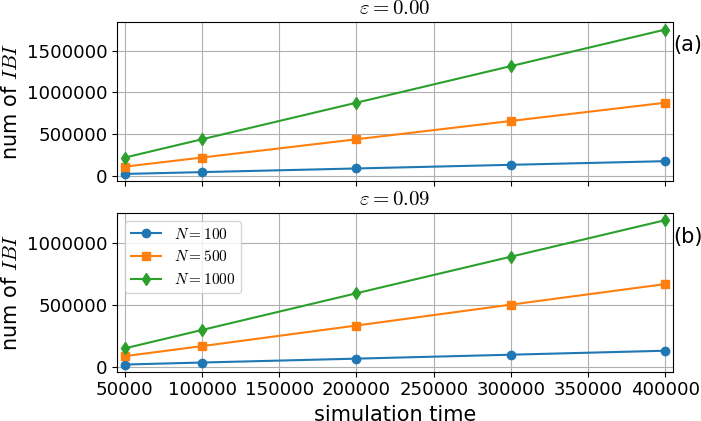}
    \caption{Number of $IBI$s as a function of simulation time for networks of $N = 100$, $N = 500$, and $N = 1\,000$ neurons. The rewiring parameter is $p = 0.010$ and the coupling strength is $\varepsilon= 0$ (a), $\varepsilon = 0.09$ (b).}
    \label{num_ibi}
\end{figure}

The simulation time, $t_{\mathrm{f}}$ was adjusted to the network size, $N$, in order to obtain a similar number of events (bursts). Figure~\ref{num_ibi} shows that the number of bursts increases linearly with the simulation time and the network size. For $N = 100$, the simulation time was $t_{\mathrm{f}} = 4\,100\,000$, for $N = 500$, $t_{\mathrm{f}} = 900\,000$, and for $N = 1000$, $t_{\mathrm{f}} = 500\,000$. In this way we had more than $10^6$ $IBI$s to calculate the ordinal probabilities (for $N = 1000$ the individual $IBI$ sequences contained more that $1000$ $IBI$s). A transient time in between $t_{0} = 100\,000-160\,000$ was disregarded.

Burst identification: as it can be observed in Fig.~\ref{rulkov_dynamics}, $y_i$ increases monotonically until a burst starts and then decreases during the burst. In this stage $y_i$ depicts small maximums every time the fast variable spikes. To appropriately detect the bursts' start times, these small maximums have to be ignored. There are several ways to filter them out; in our code we have implemented a count that re-sets to zero whenever a global maximum is identified in the slow variable.

The codes to simulate the model, detect the bursts, calculate the ordinal probabilities and the Kuramoto parameter were written in C and compiled with icc compiler (version 14.0.3)~\cite{data_availability}. The distributions were generated with the C function ``rand". Each simulation had a different seed, using the C function ``srand". The adjacency matrices were generated using NetworkX package (version 2.4)~\cite{hagberg2008exploring} implemented in Python (version 3.7.4). Specifically, we used the function that generates Watts-Strogatz networks: ``watts\_strogatz\_graph".

\section{Results}\label{4}
\begin{figure}[tb]
    \centering
    \includegraphics[width=\columnwidth]{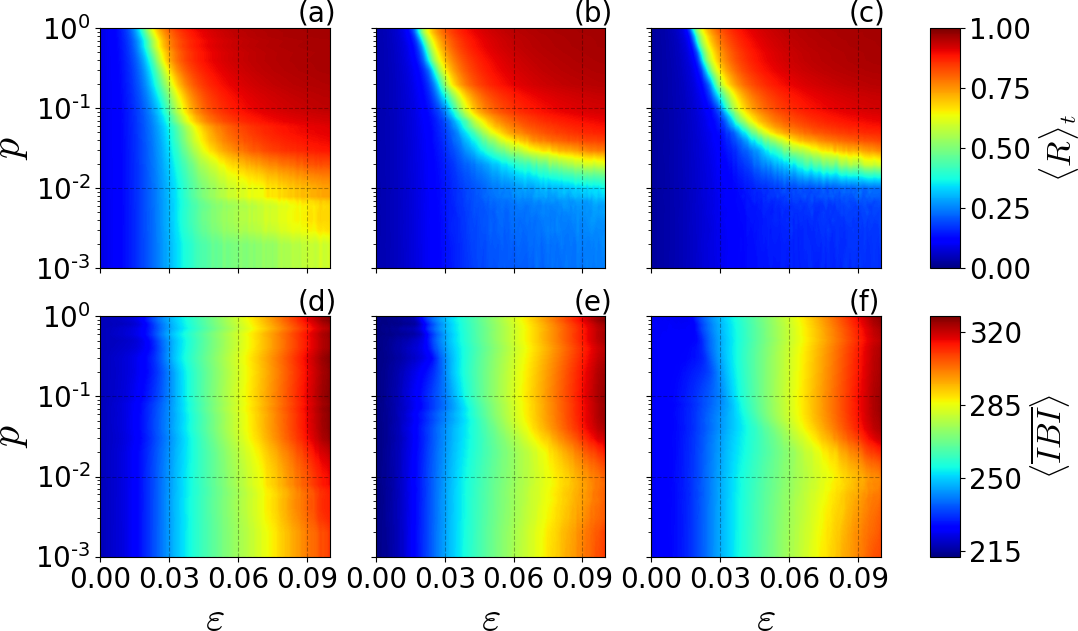}
    \caption{Mean value of the Kuramoto parameter (panels (a), (b), and (c)) and mean value of the inter-burst interval (panels (d), (e), and (f)) as a function of the rewiring probability, $p$, and the coupling strength, $\varepsilon$ for networks of $N = 100$, $N = 500$, and $N = 1\,000$ neurons, respectively.}
    \label{figure_N}
\end{figure}
The synchronization behavior is depicted in Fig.~\ref{figure_N} where the Kuramoto parameter, $\langle R \rangle_t$, is displayed in color code as a function of the coupling strength, $\varepsilon$, and the rewiring connection probability, $p$ for networks of $100$, $500$ and $ 1\,000$ neurons (panels (a)-(c) respectively). 

A transition from non-synchronized to phase-synchronized state is observed. For $p$ large enough ($p > 0.02$), the transition occurs as the coupling parameter increases \cite{wang2008synchronization,yu2011chaotic}. For $\varepsilon$ large enough ($\varepsilon > 0.025$), the increase of $p$ induces a similar synchronization transition. This behavior is observed for the three network sizes considered; however, the border is less defined when $N = 100$. When the coupling is too weak  ($\varepsilon < 0.02$) or when the network is too regular ($p < 0.02$) the transition to phase synchronization does not occur (the Kuramoto parameter is $\langle R \rangle_t < 0.6$).

Panels (d), (e), and (f) of Fig. \ref{figure_N} depict the mean value of inter-burst intervals, $\langle \overline{IBI} \rangle$, as a function of $\varepsilon$ and $p$ for $N = 100$, $500$, and $1\,000$, respectively. In the three cases, we see that the coupling strength ($\varepsilon$) increases the mean inter-burst interval. On the other hand, we note that the rewiring probability does not have a large impact in $\langle \overline{IBI}\rangle$.

\begin{figure}[tb]
    \centering
    \includegraphics[width=\columnwidth]{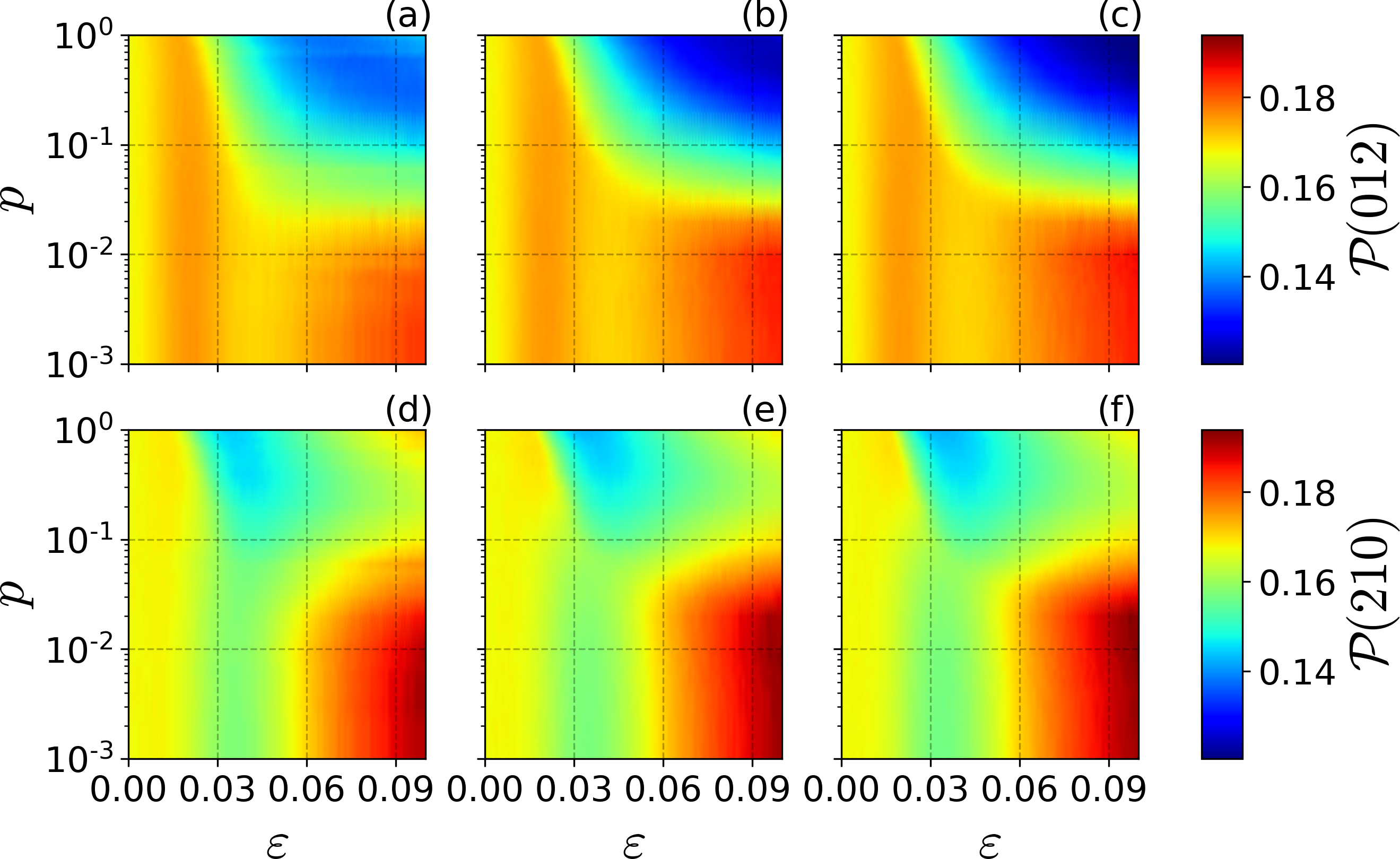}
    \caption{Probability of pattern 012 (panels (a), (b), and (c)) and of pattern 210 (panels (d), (e), and (f)) as a function of the rewiring probability, $p$, and the coupling strength, $\varepsilon$, for networks of $N = 100$, $N = 500$, and $N = 1\,000$, neurons respectively.}
    \label{symbolic_N}
\end{figure}

The results obtained with ordinal analysis are presented in Fig. \ref{symbolic_N} that depicts the probabilities of observation of pattern $012$ (panels (a), (b), and (c)) and pattern 210 (panels (d), (e), and (f)) in the parameter space ($p,\varepsilon$) considering networks of $N = 100$, $N = 500$, and $N = 1\,000$ neurons respectively. We see that $\mathcal{P}(012)$ is considerably larger (lower) than $1/6\sim 0.167$ in the regions where there is low (high) phase synchronization (see Fig. \ref{figure_N}). Thus, the probability of pattern $012$ distinguishes the regions of low and high phase synchronization.

$\mathcal{P}(210)$ (panels (d), (e), and (f)) uncovers more details, as in the low synchronization region that occurs for $p < 0.02$, it distinguishes different regions, for weak coupling ($\varepsilon < 0.03$) where pattern $210$ is under-expressed ($\mathcal{P}(210) <1/6$) and for stronger coupling ($\varepsilon > 0.60$) where 210 is over-expressed ($\mathcal{P}(210) > 0.19$). For higher $p$, a region where pattern $210$ is less expressed is also observed. This scenario is robust for the three network sizes analyzed.

\begin{figure}[tb]
    \centering
    \includegraphics[width=\columnwidth]{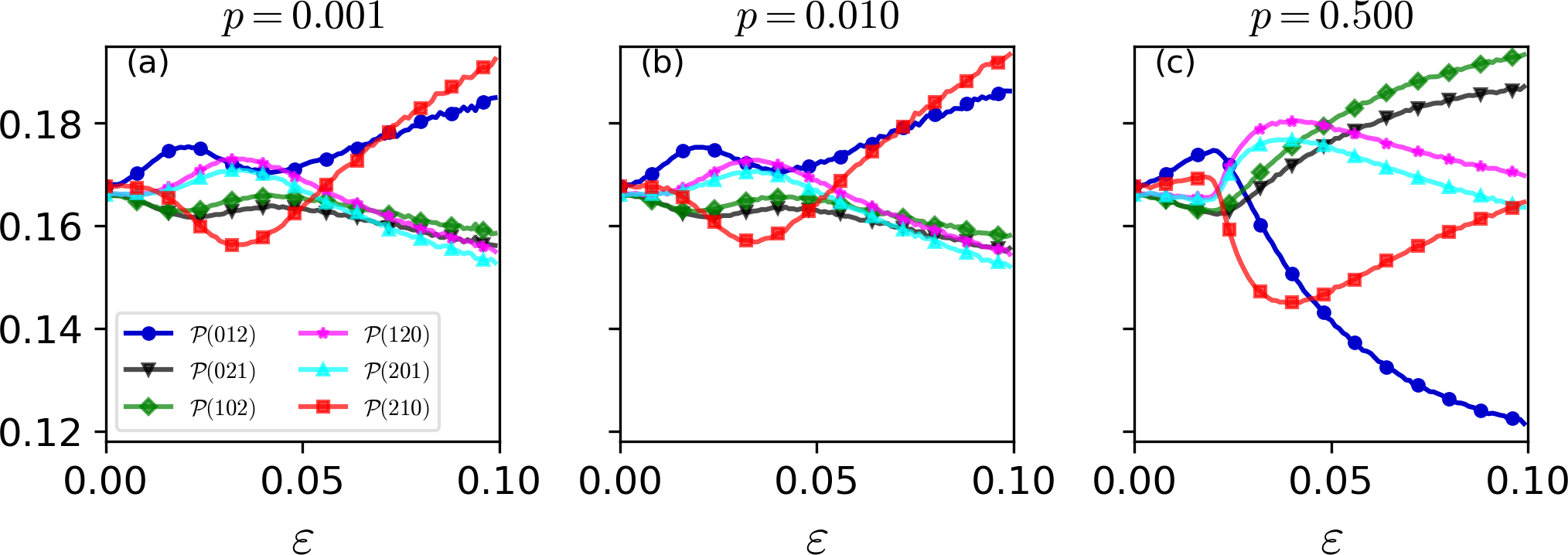}
    \caption{Probabilities of the six ordinal patterns as a function of the coupling strength when the rewiring probability is $p = 0.001$ (a), $p = 0.01$ (b), and $p = 0.5$ (c).}
    \label{fig:all_ops}
\end{figure}

\begin{figure*}[tb]
    \centering
    \includegraphics[width=1.5\columnwidth]{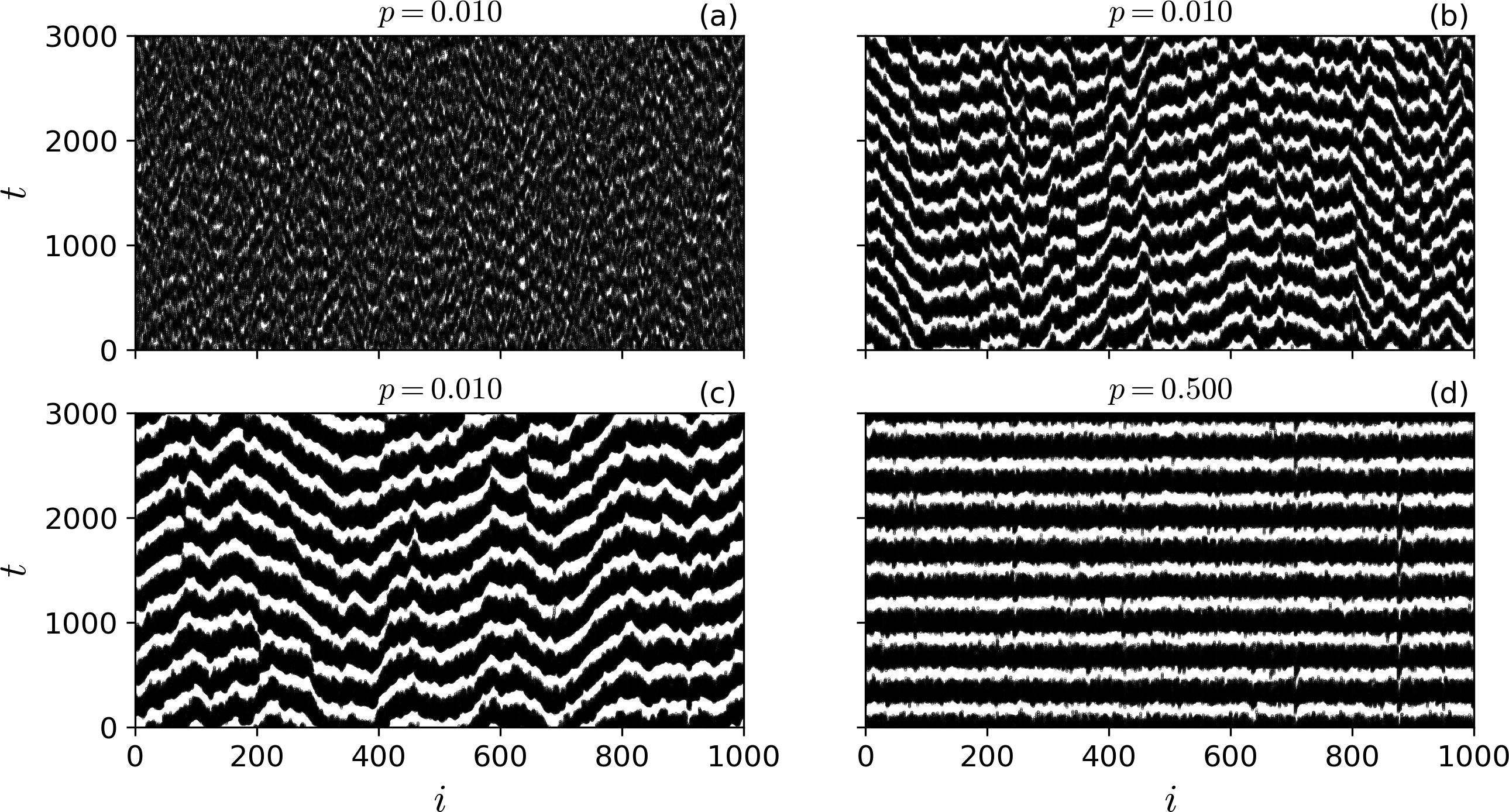}
    \caption{Raster plots representing the activity of a network of $N = 1\,000$ neurons. Panel (a) ($\varepsilon = 0.005$ and $p = 0.010$) depicts a non-synchronized state, panels (b) and (c) ($\varepsilon = 0.045$ and $p = 0.010$, and $\varepsilon = 0.090$ and $p = 0.010$, respectively) show diagonal spatio-temporal structures, and panel (d) ($\varepsilon = 0.090$ and $p = 0.500$) represents a phase-synchronized state. 
    }
    \label{raster_plot}
\end{figure*}
While the ``trend'' patterns 012 and 210 are often the most informative, the other patterns can yield interesting information as well. Figure~\ref{fig:all_ops} displays all the ordinal probabilities vs. the coupling strength, for three values of the rewiring parameter. Without coupling all the patterns have similar probabilities, consistent with the uniform distribution. Therefore, in spite of the fact that the dynamics of the uncoupled neurons is chaotic, no temporal structures are detected in the sequences of $IBI$s. Similar results are obtained when the neurons have identical parameters and are noise-free (not shown). In the presence of neuronal coupling temporal structures in the $IBI$ sequences emerge and we note that they depend on the network connectivity. For high enough coupling and low $p$ (when the network relatively regular) the ``oscillation'' patterns 021, 120, 102, and 201 are under expressed (their probabilities are $<1/6$), while for high $p$ (when the network is close to random) these patterns are over expressed (their probabilities are $>1/6$).

To investigate the reasons for the low or high values of the ordinal probabilities, we inspect the dynamics of the networks for different values of $\varepsilon$ and $p$ using raster plots, which are obtained from the fast variable of each neuron, $x_i$, considering the spike threshold of $x_{i} = 0.0$ and positive first derivative. Figure~\ref{raster_plot} depicts the raster plots for a network of $N = 1\,000$ neurons. Panel (a) for $\varepsilon = 0.005$ and $p = 0.010$ is representative of the non-synchronized state, where we see that the neural activity has low spatio-temporal coherence.
\begin{figure*}[thb]
    \centering
    \includegraphics[width=0.82\textwidth]{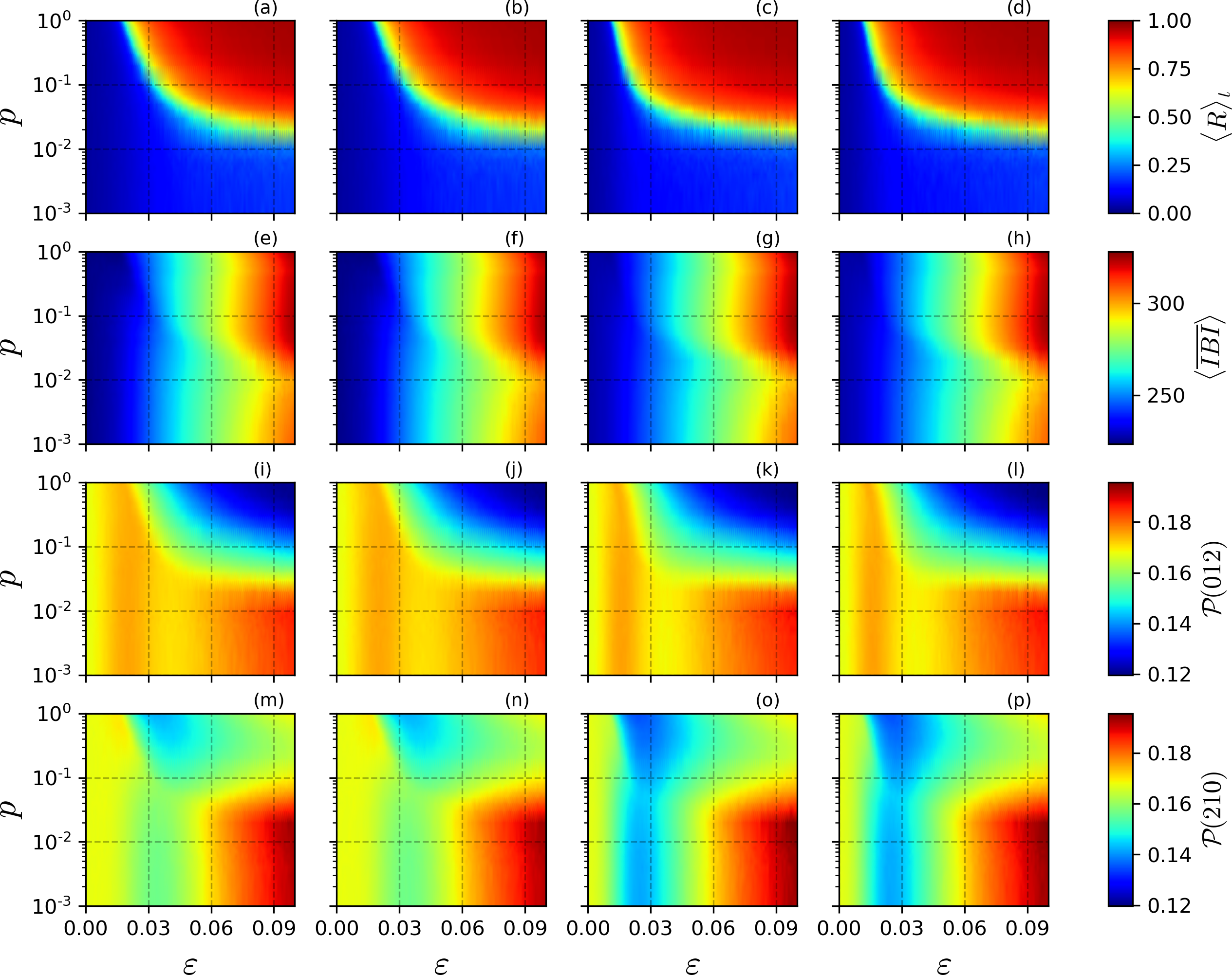}
    \caption{Mean value of the Kuramoto parameter (top row); mean $IBI$ (second row) and probabilities of patterns 012 (third row) and 210 (fourth row) for a network of $ 1\,000$ neurons, considering the four cases described in the text: noisy heterogeneous neurons (case i, left column); deterministic (noise-free) heterogeneous neurons (case ii, second column); noisy identical neurons (case iii, third column); deterministic identical neurons - (case iv, right column). } 
    \label{figure_cases}
\end{figure*}

In contrast, panels (b) and (c) ($\varepsilon = 0.045$ and $p = 0.010$, and $\varepsilon = 0.090$ and $p = 0.010$, respectively) show a different situation, in which diagonal structures are observed, which resemble zig-zag fronts \cite{wang2008synchronization,budzinski2019synchronous,osipov2005synchronized,ivanchenko2007network}. In this situation, the network can depict groups of neurons with different mean frequencies \cite{budzinski2019synchronous}. The presence of horizontal structures in these plots would indicate that most of neurons start their bursts at similar times, that means, that they are phase synchronized. However, horizontal structures are almost absent in Figs.~\ref{raster_plot}(b) and (c). Because the Kuramoto order parameter quantifies the level of phase synchronization in the network, the absence of horizontal structures explains why the Kuramoto parameter has a low value (for the parameters of Figs.~\ref{raster_plot}(b) and (c), $\langle R \rangle_t < 0.6$).

Figure~\ref{raster_plot}(d) for $\varepsilon = 0.09$ and $p = 0.5$ is representative of the phase-synchronized states, in which the horizontal structures are observed and $\langle R \rangle_t$ is high. These structures show that the bursts begin at similar times. 
 
An important question is the role of noise and heterogeneous neurons' parameters. So far we have considered $\alpha_{i}$ and $I_{i,\mathrm{noise}}$ Gaussian distributed. To determine how they affect the dynamics, we simulate $ 1\,000$ neurons and compare four cases:

     (i) heterogeneous neurons with noise ($\alpha_{i}$ and $I_{i,\mathrm{noise}}$ distributed as indicated in Sec.~ \ref{implementation});
     
     (ii) heterogeneous neurons without noise ($\alpha_{i}$ distributed as indicated above, $I_{i,\mathrm{noise}}=0$);
     
    (iii) identical neurons with noise ($\alpha_{i}=4.25$ and $I_{i,\mathrm{noise}}$ distributed as indicated in Sec.~\ref{implementation});
    
    (iv) identical neurons without noise ($\alpha_{i}=4.25$, $I_{i,\mathrm{noise}}=0$).
%
 
Figure~\ref{figure_cases} depicts the results. Here, panels (a), (b), (c), and (d) depict the mean value of Kuramoto parameter, $\langle R \rangle_t$, as a function of $\varepsilon$ and $p$ for the cases (i), (ii), (iii), and (iv), respectively. A very similar behavior is observed for all cases.

The second and third rows in Fig. \ref{figure_cases} depict the mean value of inter-burst intervals, $\langle \overline{IBI}\rangle$, and the probability of pattern 012 respectively, and we again see a very similar behavior in the four cases cases. In contrast, the probability of pattern 210 (fourth row) uncovers differences between the dynamics of the deterministic (noise free) network and the dynamics of the stochastic network. For intermediate coupling ($ 0.01 < \varepsilon < 0.04$), without noise $\mathcal{P}(210)$ takes lower values  ($< 0.16$) than with noise. 

In the four cases considered we see that the probability of pattern 210 distinguishes, in the region $p < 0.020$ (almost regular networks), different dynamical behavior depending on the coupling strength. In contrast, $\langle R \rangle_t$ only characterizes the dynamics as non-synchronized states.

\begin{figure}[tb]
    \centering
    \includegraphics[width=\columnwidth]{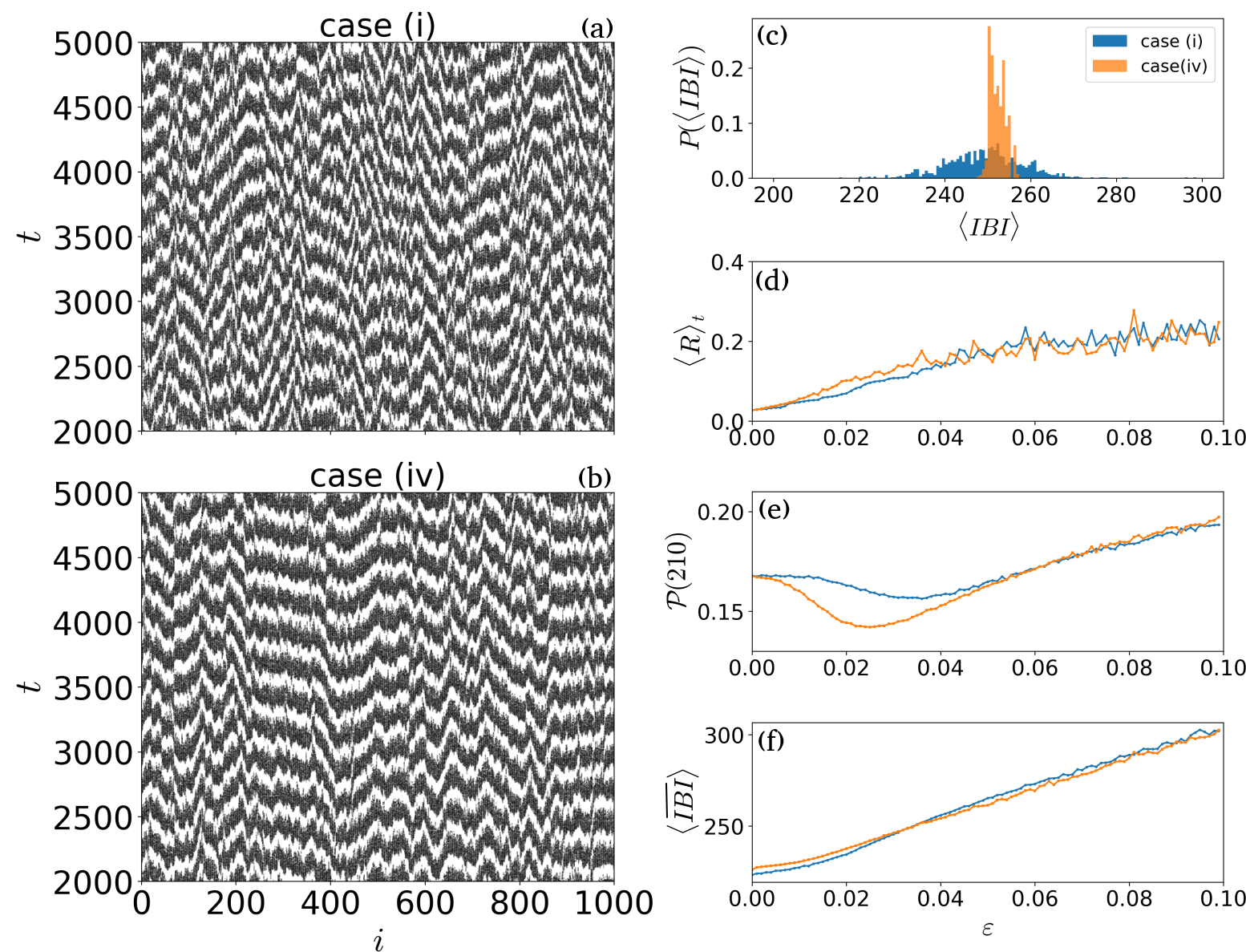}
    \caption{(a), (b) Raster plots for a network of $N = 1\,000$ neurons, when the coupling strength is $\varepsilon = 0.025$ and the rewiring parameter is $p = 0.010$. In (a) the neurons are heterogeneous and stochastic (case (i)); in (b), they are noise-free and identical (case (iv)). Panel (c) shows the distribution of mean $IBI$s of the individual neurons, $\langle IBI_i \rangle$. We see that case (iv) depicts a more localized distribution. Panels (d), (e), and (f) show the $\langle R \rangle$, $\mathcal{P}(210)$, and $\langle \overline{IBI}\rangle$, respectively, for $p = 0.010$ as a function of $\varepsilon$.}
    \label{raster_plot_2}
\end{figure}

To investigate the origin of the differences found in $\mathcal{P}(210)$ in the region ($p < 0.1$ and $0.01 < \varepsilon < 0.04$), we analyze the spatio-temporal dynamics of the network. Figure~\ref{raster_plot_2} depicts the raster plot for $\varepsilon = 0.025$, $p = 0.01$ when the neurons are heterogeneous and noisy (case i, panel a), and when they are identical and noise-free (case iv, panel b).  We see a similar partially coherent spatio-dynamics; however, we can also notice subtle differences: panel (a) depicts less coherent spatial structures than panel (b), where more partially horizontal structures are observed. The similar dynamics is captured by the Kuramoto parameter that takes similar values, $\langle R \rangle_t = 0.092$ in (a) and  $\langle R \rangle_t = 0.098$ in (b). The subtle differences might be the reason why $\mathcal{P}(210)$ is different in the two cases: $\mathcal{P}(210) = 0.16$ in (a); $\mathcal{P}(210) = 0.14$ in (b).

To compare the temporal dynamics, panel (c) depicts the distribution of the inter-burst intervals of the individual neurons, averaged in time, $\langle {IBI}_i \rangle=\langle IBI_{\mathrm{k},i}\rangle_{\mathrm{k}}$. We observe a larger dispersion of the distribution in case (i) than in case (iv). Panels (d), (e), and (f) depict $\langle R \rangle_t$, $\mathcal{P}(210)$ and  $\langle \overline{IBI} \rangle$ respectively, as a function $\varepsilon$ for the two cases. Despite the similar values of $\langle R \rangle_t$ and $\langle \overline{IBI} \rangle$, $\mathcal{P}(210)$ shows, in a range of coupling strengths, different values for cases (i) and (iv). 

\begin{figure}[t]
    \centering
    \includegraphics[width=\columnwidth]{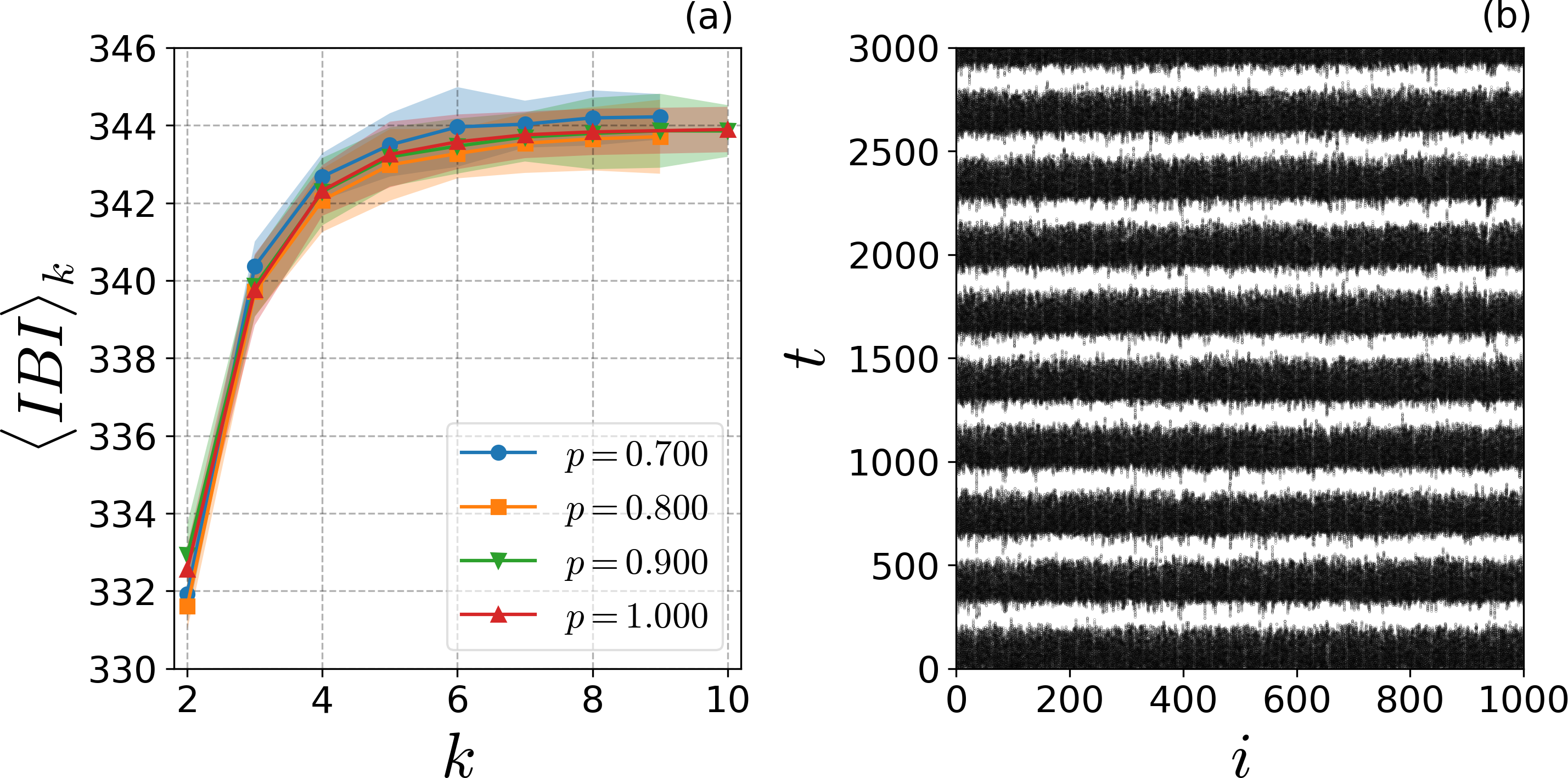}
    \caption{Panel (a) shows the average degree of a neuron $\langle \overline{IBI}_i\rangle$ as a function of its degree, $k_i$, for  a network of $N = 1\,000$ heterogeneous neurons with noise (case (i)),  $\varepsilon = 0.08$ and different values of $p$. Results are obtained from $10$ simulations and the dispersion is represented by the filled area. Panel (b) depicts the raster plot for $\varepsilon = 0.08$ and $p = 0.8$. In this situation, the network is phase synchronized and $\langle R \rangle = 0.958$.}
    \label{permutation_entropy}
\end{figure}

To complement the analysis done at the macroscopic level of the global network, we now characterize the microscopic dynamics by analyzing the burst sequences of the individual neurons. Our goal is to determine if there is a statistical relation between the way a neuron is connected and the properties of its sequence of bursts. Specifically, we investigate how the number of links (i.e., the degree $k$ of a neuron) affects the $IBI$ sequence. In networks of coupled oscillators it was recently found, numerically and experimentally ~\cite{tlaie2019dynamical,tlaie2019statistical}, that there is a relation between the dynamical complexity and the degree: nodes with higher degree were found to have lower levels of complexity. 

Considering a network of $1000$ noisy heterogeneous neurons (case i), Fig.~\ref{permutation_entropy}(a) shows that the mean $IBI$ of a neuron, $\langle IBI \rangle_{i}$, increases with the degree of the neuron, $k_i$. In addition, the standard deviation of $IBI$ distribution of each neuron
decreases with the degree (not shown) indicating that neurons that have several connections have a more regular bursting activity than neurons that have few connections.

It is interesting to note that, for the parameters used in Fig.~\ref{permutation_entropy}(a), the Kuramoto parameter has a large value, $\langle R \rangle_t>0.95$, however, the mean $IBI$ is not the same for all the neurons: it increases with the neuron's degree.  The raster plot, shown in Fig.~\ref{permutation_entropy}(b), confirms that the neurons are synchronized in phase (horizontal structures are clearly visible), but the synchronization is not perfect. While in this plot the irregular fluctuations appear to be random, our analysis of the $IBI$ distribution of each neuron (the fact that the mean value increases and the standard deviation decreases with the neuron's degree) indicates that fluctuations are not only due to the added noise but have a dynamical origin, and in fact, similar results are found in identical and noise-free neurons (not shown).

\section{Discussion}\label{5}

Through extensive simulations we have shown that ordinal analysis is a useful tool to analyze neuronal activity. We have shown that it can uncover subtle differences in the spatio-temporal dynamics, which are not seen by other indicators (such as the mean inter-burst-interval, or the Kuramoto parameter). The main advantages of this methodology are that it can be applied to raw, unprocessed data and is rather unaffected by the presence of noise, outliers or missing data~\cite{bandt}. In spite of having been extensively used to analyze biomedical  signals (EEG, ECG, etc.~\cite{parlitz2012classifying,zanin2012permutation,carlos}), few studies have applied ordinal analysis to neuronal spike trains. The data requirement is a main limitation, as (to the best of our knowledge) simultaneous intra-cellular recordings of the membrane potential of a large number of cells, during a period of time long enough to record 100s or 1000s of spikes, are not yet freely available. Therefore, efforts so far have focused in simulated neuronal spikes.

Another limitation of the ordinal methodology is that it does not consider amplitude information, i.e., the actual values of the data points are disregarded. Different generalizations have been proposed in order to also take into account amplitude information~\cite{wop,aape,antonio}. Because in neuroscience the standard way to characterize spike trains is through the analysis of the distribution of the inter-spike intervals, here we have complemented the results obtained with ordinal analysis, with the analysis of the $IBI$ distribution, both, at the microscopic level of the individual neurons and at the macroscopic level of the global network.
    
In our network we have not detected any relation between the permutation entropy (a complexity measure~\cite{bandt2002permutation}) of the $IBI$ sequence of a neuron and the degree of the neuron: in the parameter regions analyzed, $S_i$ was found to be almost independent of $k_i$ (not shown). However, we have found parameter regions in which the neurons with the largest number of connections display a more regular bursting activity (the mean value of the $IBI$ distribution increases and its standard deviation decreases with the number of links a neuron has). This can be interpreted as due to the fact that neurons with large degree receive stronger coupling with respect to those with low degree. We remark that in the model equations the coupling term is normalized to the average degree, therefore, this term is stronger in neurons that have several links with respect to neurons with few links. The fact that the coupling increases the time between consecutive bursts can be seen at, the macroscopic level, in the second rows of Figs.~\ref{figure_N} and \ref{figure_cases}. It is left for future work to use more advanced complexity measures~\cite{tlaie2019dynamical,tlaie2019statistical} to further characterize how the complexity of the neuronal dynamics depends on the global connectivity and on the individual activity of the neurons (excitable firing, periodic firing or bursting).

The degree of synchronization has been quantified with the time-averaged Kuramoto parameter, $\langle R \rangle_t$; however, it only provides partial information about the synchronization of the phases of the neurons' bursts. We have found parameter regions where $\langle R \rangle_t$ takes a high value, but the neurons have different bursting frequencies (we have shown that the average $IBI$ of an individual neuron depends on the neuron's degree). For future work, it would be interesting to characterize the synchronized activity using other measures~\cite{peter,ernesto}, in particular, the ordinal synchronization measure proposed in~\cite{javier}, or the mutual information of time series of ordinal patterns proposed in~\cite{maria_preprint}.

The neuronal coupling considered here is certainly not realistic, as it is linear and constant in time. The networks considered have a low density of links (four links per neuron). As future work, it would be very interesting to extend this study to other types of coupling (pulsed, excitable or inhibitory, that change in time, etc.), to consider denser networks and to consider structured networks (with layers or modular structure, with hubs and/or dead-end nodes, etc.).

\section{Conclusions}\label{6}

We have used symbolic ordinal analysis to investigate the dynamics of an ensemble of Rulkov neurons  mutually coupled in a Watts-Strogatz network. We have considered parameters such that the individual neurons fire bursts of spikes, and we have characterized the sequences of the inter-burst-intervals ($IBI$s) by computing the probabilities of the ordinal patterns. We have found that these ordinal probabilities allow to identify different dynamical regimes, which depend on the coupling strength and the network topology. These regimes are not differentiated by the average $IBI$ or by the Kuramoto order parameter; however, different spatio-temporal structures were seen in the raster plots (such as non-synchronized states, phase-synchronized ones, and zig-zag structures). We have shown that our results are valid for different network sizes and are robust to the presence of noise and heterogeneous neurons' parameters.

\section*{Acknowledgments}

This study was financed in part by the Coordena\c c\~ao de Aperfeiçoamento de Pessoal de N\'{\i}vel Superior - Brasil (CAPES) - Finance Code 001, Conselho Nacional de Desenvolvimento Cient\'{\i}fico e Tecnol\'ogico,  CNPq - Brazil, grant number 302785/2017-5, and Finan\-ciadora de Estudos e Projetos (FINEP). Computer simulations were performed at the LCPAD cluster at Universidade Federal do Paran\'{a} (FINEP - CTINFRA). C. Masoller acknowledges partial support from Spanish Ministerio de Ciencia, Innovaci\'{o}n y Universidades grant PGC2018-099443-B-I00 and ICREA ACADEMIA, Generalitat de Catalunya. C. Masoller also acknowledges the hospitality of the Universidade Federal do Paran\'{a}, where a collaboration was established and this work started.

\bibliographystyle{model1-num-names.bst}

\end{document}